\documentclass[11pt]{article}
\usepackage{fa2023}
\usepackage{amsmath}
\usepackage{cite}
\usepackage{url}
\usepackage{graphicx}
\usepackage{color}
\usepackage{siunitx}
\usepackage[utf8]{inputenc}

\usepackage{amssymb}
\usepackage[caption=false,font=footnotesize]{subfig}
\usepackage[export]{adjustbox}
\usepackage[super]{nth}

\DeclareMathOperator*{\softmax}{soft-max}

\DeclareMathOperator{\MultiHead}{MultiHead}
\DeclareMathOperator{\Concat}{Concat}
\DeclareMathOperator{\Attention}{Attention}

\title{Permutation Invariant Recurrent Neural Networks for Sound Source Tracking Applications}

\multauthor
{David Diaz-Guerra$^{1*}$ \hspace{1cm} Archontis Politis$^1$ \hspace{1cm} Antonio Miguel$^2$} { \bfseries{Jose R. Beltran$^2$ \hspace{1cm} Tuomas Virtanen$^1$}\\
  $^1$ Audio Research Group, Tampere University, Finland\\
$^2$ Department of Electronic Engineering and Communications, University of Zaragoza, Spain\\
\correspondingauthor{david.diaz-guerra@tuni.fi}{David Diaz-Guerra et al.}
}

\begin{document}

\maketitle
\begin{abstract}
Many multi-source localization and tracking models based on neural networks use one or several recurrent layers at their final stages to track the movement of the sources. Conventional recurrent neural networks (RNNs), such as the long short-term memories (LSTMs) or the gated recurrent units (GRUs), take a vector as their input and use another vector to store their state. However, this approach results in the information from all the sources being contained in a single ordered vector, which is not optimal for permutation-invariant problems such as multi-source tracking. In this paper, we present a new recurrent architecture that uses unordered sets to represent both its input and its state and that is invariant to the permutations of the input set and equivariant to the permutations of the state set. Hence, the information of every sound source is represented in an individual embedding and the new estimates are assigned to the tracked trajectories regardless of their order.
\end{abstract}

\section{Introduction}\label{sec:introduction}

In recent years, the state-of-the-art of sound source localization established by classic signal processing techniques has been surpassed by new systems using deep-learning models \cite{grumiaux2022}. These models use different input features and network architectures, but most of them track the temporal evolution of the signals using convolutional layers followed by recurrent layers \cite{adavanne2019, perotin2019a, cao2019}. Using these architectures, the latent representations at every hidden layer are difficult to interpret and we cannot exploit the permutation invariance of the tracking problem where, if we cannot apply any criteria to order or classify the sources, any permutation of the sources should be considered equally correct.

In \cite{diaz-guerra2022c}, we proposed an icosahedral convolutional neural network (icoCNN) for single source localization where the output of the last convolutional layer can be interpreted as the probability distribution of the direction of arrival (DOA) and we can obtain the estimated DOA as its expected value. Extending this model to multi-source scenarios is straightforward and we just need to increase the number of channels of the last convolutional layers to the maximum number of concurrent sources $M$ that the model should be able to localize. Following this approach, after computing the expected value of every one of the $M$ probability distributions generated by the icoCNN, we obtain a set of $M$ DOAs that should be considered invariant to the permutations of its elements. In order to incorporate a recurrent neural network (RNN) after the localization model to increase its temporal perceptive field and improve its tracking capabilities, we could concatenate every element of the DOA set into a single vector and use it as the input of a gated recurrent unit (GRU) \cite{cho2014} or a long short-term memory (LSTM) layer \cite{hochreiter1997a}. However, we should expect the output of a tracking system to not be affected by the order of the new estimates at every time frame (i.e., to be invariant to the permutations of the input set), and a conventional RNN operating over the concatenation of the estimates would need to learn this property during the training instead of being part of its architecture. In addition, in a tracking system, we can also expect the association of a new estimate to the tracked trajectories be done regardless of their order (i.e., be equivariant to the permutations of the state set) but the state vector generated by a conventional RNN would contain the information of every tracked trajectory in an unstructured way so we would not be able to exploit this property either.

In this paper we present a permutation-invariant recurrent neural network (PI-RNN) that takes an unordered set of embeddings as input (each one with the information of one of the sources detected by the localization network) and generates a recursive output, or state, that is also an unordered set of embeddings with the information of every tracked trajectory. As we could expect from a tracking system, the proposed architecture associates the embeddings in the input set to the embeddings of the state set in a way that is invariant to the permutations of the input set and equivariant to the permutations of the state set. 

To the best of our knowledge, this is the first recurrent layer that works with sets instead of with vectors. The closest proposal in the literature is probably the TrackFormer \cite{meinhardt2022}, a model for multiple object tracking on video signals that is based on the DETR transformer \cite{carion2020, zhu2022}, a model for object detection on images. The recursivity of the TrackFormer model is built around the decoder of the DETR transformer by using the output obtained for a video frame as the input for the following frame. Compared with the TrackFormer, the PI-RNN is not a model but a layer that can be integrated easily into many different models. In addition, it is based on an architecture, the conventional GRU, that, unlike the transformer, was designed to be used in recurrent loops. 

Thanks to be taking into account the symmetries of the problem, the proposed PI-RNN, compared with the conventional RNN, scales better with the number of tracked sources and the amount of information stored in each one. Furthermore, we present experiments proving that they can obtain better tracking results than the conventional GRUs.
\section{Network architecture}

\begin{figure}[tb]
    \centering
    \includegraphics[width=0.95\linewidth]{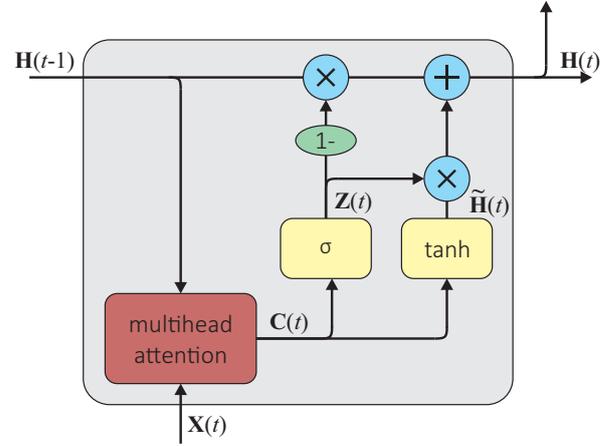}
    \caption{Architecture of the proposed permutation invariant recurrent layer.}
    \label{fig:PI-GRU}
\end{figure}

Conventional RNNs use a $\textbf{h}(t) \in \mathbb{R}^{d_h}$ vector to store the tracking state, which is updated at every time frame based on an input vector $\textbf{x}(t) \in \mathbb{R}^{d_x}$ using fully connected perceptrons whose computational complexity and number of trainable parameters grow linearly with $d_x$ and quadratically with $d_h$. When applied to track up to $M$ sources, the information of all the sources and tracked trajectories are stored in these vectors without any structure, so there is a trade-off between the number of sources $M$ that we can track, the amount of information that we store about each one, and the model size and complexity. 

In contrast to conventional RNNs, we propose to replace the input and state vectors $\textbf{x}(t)$ and $\textbf{h}(t)$ with the sets of embedding $\textbf{X}(t) = \{ \textbf{x}_1(t), \textbf{x}_2(t), ..., \textbf{x}_{M_X}(t) \}$ and $\textbf{H}(t) = \{ \textbf{h}_1(t), \textbf{h}_2(t), ..., \textbf{h}_{M_H}(t) \}$ where every element $\textbf{x}_i(t) \in \mathbb{R}^{d_x}$ and $\textbf{h}_i(t) \in \mathbb{R}^{d_h}$ contains information about a single input detection or tracked trajectory respectively. For the sake of simplicity, we will keep $M_X=M_H=M$ and $d_x=d_h=d$ during the rest of the paper, however the proposed architecture can work with $M_X \neq M_H$ or even with dynamic values that change during time, and it can be easily extended to configurations with $d_x \neq d_h$.

In order to match every new embedding of the input set with the embeddings of the state set, we can use a multi-head attention module \cite{vaswani2017}, which is well known for its use in transformer models and is invariant to the permutation of the elements of its input sets:
\begin{equation}
\begin{split}
	\mathbf{C}(t) = \MultiHead(& \mathbf{H}(t-1),\  \\
 & \mathbf{X}(t)\cup\mathbf{H}(t-1), \\
 & \mathbf{X}(t)\cup\mathbf{H}(t-1))
\end{split}
\end{equation}
\begin{equation}
	\label{eq:multihead}
    \MultiHead(\mathbf{Q},\mathbf{K},\mathbf{V}) = \Concat(\mathbf{head}_1, ..., \mathbf{head}_{N_{\textrm{heads}}})
\end{equation}
\begin{equation}
    \mathbf{head}_i = \Attention(\mathbf{Q}\mathbf{W}^Q_i, \mathbf{K}\mathbf{W}^K_i, \mathbf{V}\mathbf{W}^V_i)
\end{equation}
\begin{equation}
	\label{eq:attention}
     \Attention(\mathbf{Q}_i, \mathbf{K}_i, \mathbf{V}_i) = \softmax\left(\frac{\mathbf{Q}_i\mathbf{K}_i^T}{\sqrt{d_k}}\right)\mathbf{V}_i,
\end{equation}
with the $\softmax(\cdot)$ operating across rows.
With this configuration, the generated set $\mathbf{C}(t)$ is invariant to the permutations of $\mathbf{X}(t)$ and equivariant to the permutations of $\mathbf{H}(t-1)$ as we would expect from a tracking system.

\begin{figure}[tb]
    \centering
    \includegraphics[width=0.95\linewidth]{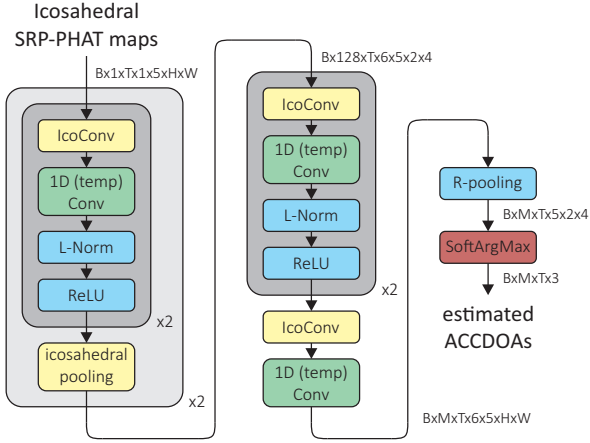}
    \caption{Architecture of the icoCNN used for evaluation. B is the batch size, T is the number of temporal frames of the acoustic scenes, $H=2^r=8$ and $W=2^{r+1}=16$ are the height and the width of the projections of the icosahedral grid.}
    \label{fig:icoCNN}
\end{figure}

Finally, as shown in Fig. \ref{fig:PI-GRU}, once we have assigned the input embeddings to their corresponding state embedding, we can just update every element of the state set according to this assignation:
\begin{equation}
	\mathbf{h}_i(t) = \left[1-\mathbf{z}_i(t)\right] \odot \mathbf{h}_i(t-1) + \tilde{\mathbf{h}}_i(t)
\end{equation}
\begin{equation}
	\mathbf{z}_i(t) = \sigma(\mathbf{c}_i(t)\mathbf{W}^z)
\end{equation}
\begin{equation}
     \tilde{\mathbf{h}}_i(t) = \tanh(\mathbf{c}_i(t)\mathbf{W}^h),
\end{equation}
where $\odot$ denotes element-wise vector multiplication and $\sigma(\cdot), \tanh(\cdot)$ denote sigmoid and hyperbolic tangent functions respectively, applied to each element of their vector arguments.
This gated architecture is based on a simplified version of the minimal gated recurrent unit \cite{zhou2016a}, but we could design different architectures based on different conventional recurrent architectures. As in conventional RNNs, the number of trainable parameters grows quadratically with $d$, but in the case of the PI-GRUs we have $M$ embeddings of size $d$ containing the information of every tracked trajectory. Hence, we can expect our model to scale better when we increase the number of sources we want to track or the amount of information that we want to be able to represent for each one of them.

\section{Evaluation}

\subsection{Experiment design}

\begin{figure}[tb]
    \centering
    \includegraphics[width=0.95\linewidth]{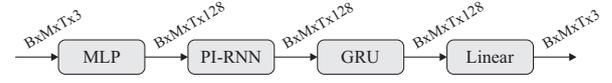}
    \caption{Architecture of the PI-RNN used after the icoCNN in the evaluated model.}
    \label{fig:PI-RNN}
\end{figure}

As a preliminary study of the performance of this new architecture, we decided to add a PI-RNN after the icoCNN presented in \cite{diaz-guerra2022c} for single source localization. As shown in Fig. \ref{fig:icoCNN}, in order to extend the icoCNN to multi-source localization, we just increased the number of output channels from $1$ to $M$. Fig. \ref{fig:PI-RNN} represents the PI-RNN we used after the icoCNN: we first used a multi-layer perceptron to project every ACCDOA \cite{shimada2021} generated by the icoCNN into an embedding of size $d$ and then we used those embeddings as the input set of our PI-RNN. After the PI-RNN had associated every new estimate from the icoCNN to one of the tracked trajectories, we added a conventional GRU (operating independently over the embedding of every tracked trajectory so it did not break the permutation invariance of the model) and, finally, we used a linear layer to project the $d$-size embedding into a 3D ACCDOA. The initial state of every embedding of the state set of the PI-RNN was learned during the training of the model while, at every time frame, the embeddings of all the inactive trajectories were reset (i.e., those who had lead to ACCDOAs with a norm lower than 0.5).

\begin{figure}[t]
    \centering
    \includegraphics[width=0.95\linewidth]{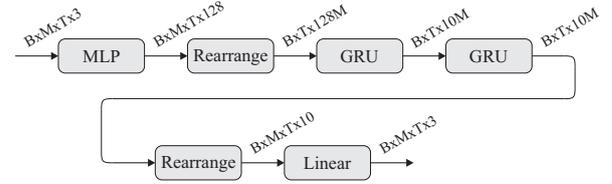}
    \caption{Architecture of the conventional RNN used after the icoCNN in the baseline model.}
    \label{fig:RNN}
\end{figure}

\begin{figure*}[tb]
    \centering
    \subfloat{
        \includegraphics[width=0.25\linewidth,valign=c]{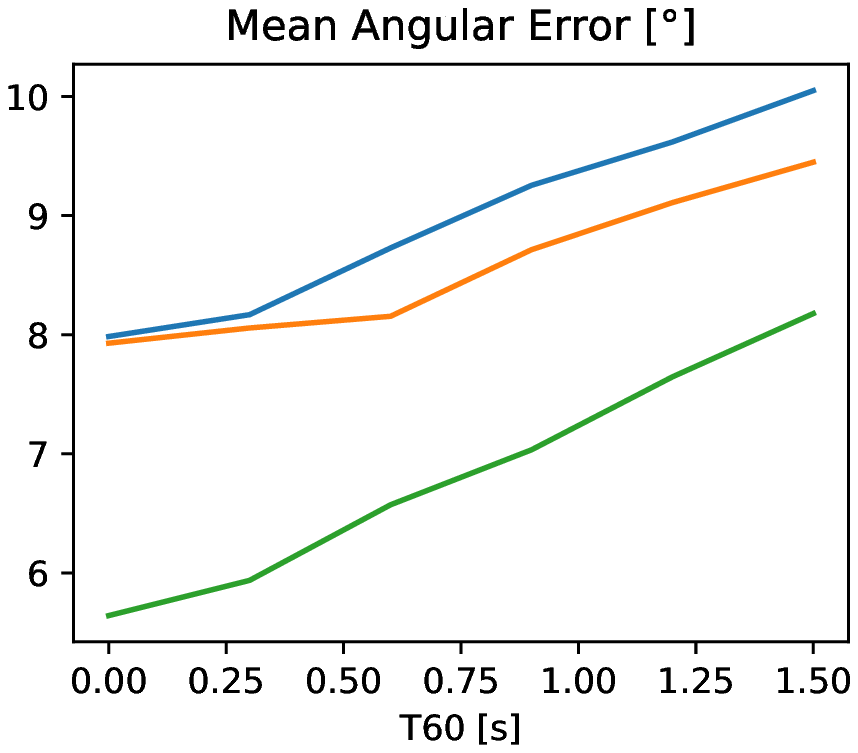}
    }
    \subfloat{
        \includegraphics[width=0.26\linewidth,valign=c]{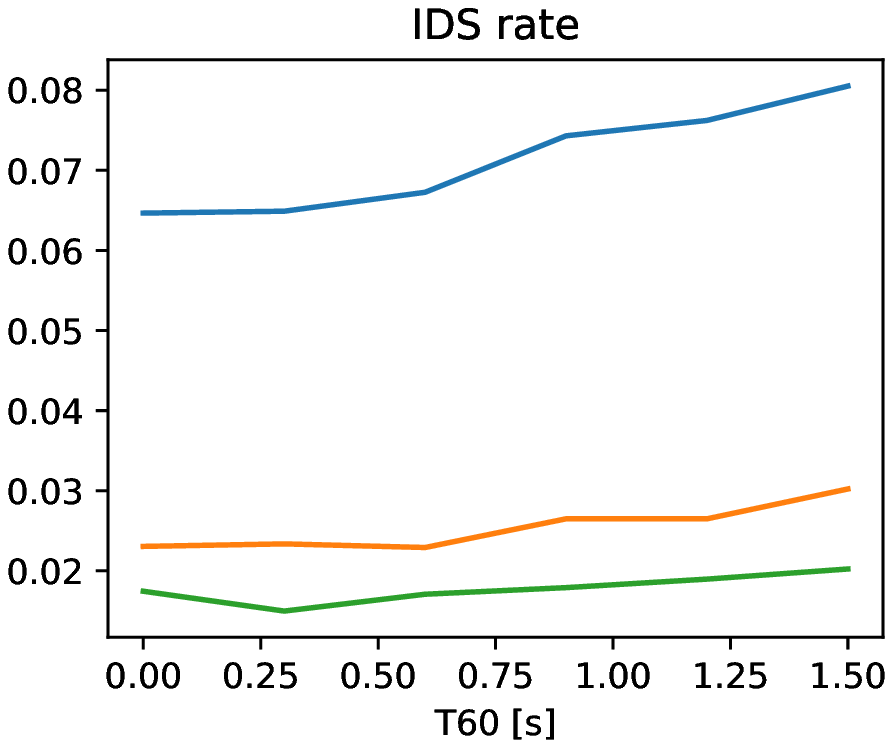}
    }
    \subfloat{
        \includegraphics[width=0.28\linewidth,valign=c]{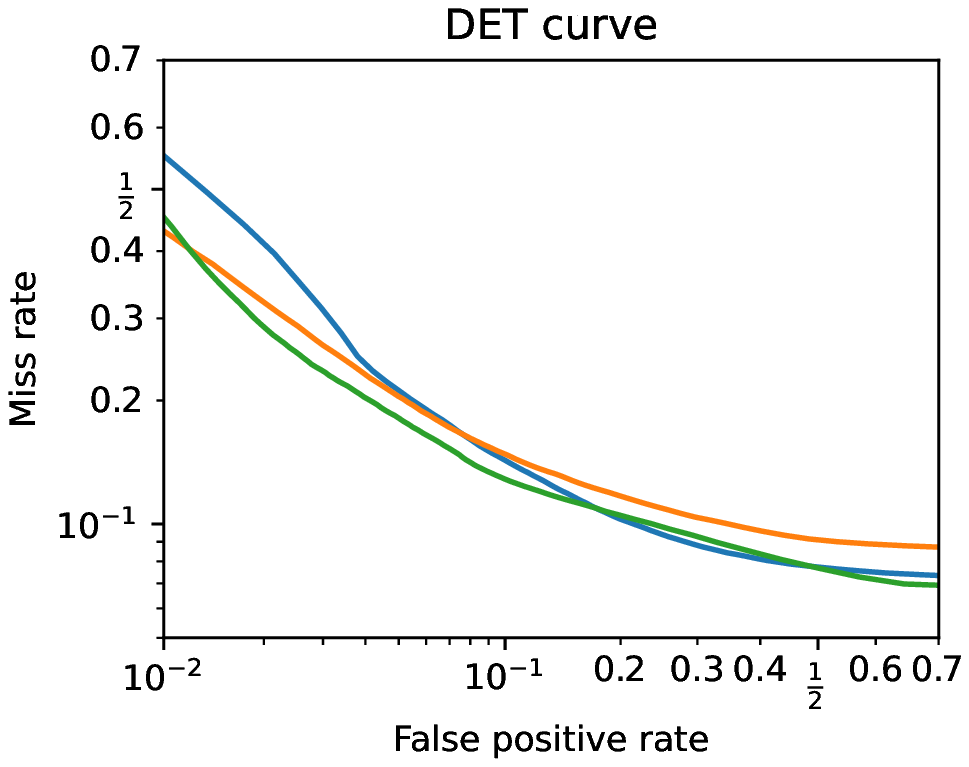}
    }
    \subfloat{
        \includegraphics[width=0.17\linewidth,valign=c]{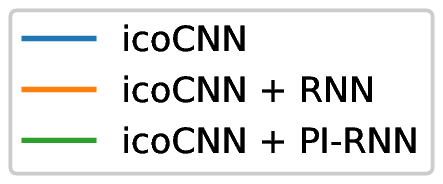}
    }\hfill
    \caption{Evaluation metrics for proposed PI-RNN and the baseline models.}
    \label{fig:results}
\end{figure*}

\begin{figure}[tb]
    \centering
    \includegraphics[width=0.95\linewidth]{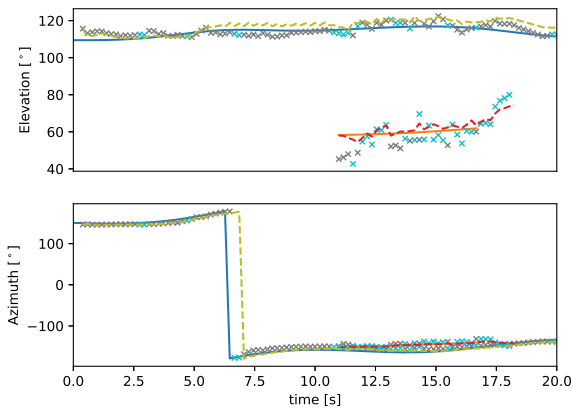}
    \caption{Example of one of the test acoustic scenes. The solid line represents the ground truth trajectories of the sources, the crosses the defections estimated at the icoCNN (i.e., the input of the PI-RNN), and the dashed line the trajectories estimated by the whole model (i.e., the output of the PI-RNN). The color indicates to which of the outputs they correspond, so the IDS are visible.}
    \label{fig:example}
\end{figure}

The method was compared to two baselines, a) the icoCNN without any kind of recurrent layers, and b) the icoCNN with two conventional GRUs designed to have a similar number of trainable parameters as the evaluated model (see Fig. \ref{fig:RNN}). In order to avoid identity switches (IDSs) in the tracked trajectories, we trained all the models using sliding permutation invariant training (sPIT) \cite{diaz-guerra_position_2022}. To facilitate the training of the icoCNN, we added an auxiliary frame-level permutation invariant training (fPIT) at its output in the models that included recurrent layers after it.

We used the same synthetic dataset as in \cite{diaz-guerra_position_2022}, where acoustic sources randomly appeared and disappeared along 20-second-length scenes. As source signals, we used speech utterances from the LibriSpeech corpus and we simulated them following random trajectories in rooms with reverberation times from $T_{60}=$ 0.2 to \SI{1.3}{\s} with the image source method. The maximum number of concurrent active sources in a time frame was 3.

We used $M=10$ as the number of ACCDOA outputs of all our models since we observed that it was beneficial to use a higher number than the maximum possible number of active sources in the dataset (i.e., 3) and we used $d=128$ as embedding size for the input and state sets of the PI-RNN. This is a preliminary study of this new architecture and further experiments should be conducted for a better optimization of these hyperparameters.

\subsection{Results}

As we can see in Fig. \ref{fig:results}, the proposed PI-RNN clearly outperforms the baselines in terms of localization error and the frequency of the identity switches while, as we can see in the detection error tradeoff (DET) curve, the trade-off between false positives and misses remains is for all the evaluated models. It is worth saying that both the conventional and the permutation-invariant RNNs are receiving only spatial information about the estimated sources. By modifying the model to include spectral information in their input we could expect both models to improve their performance, with the PI-RNNs scaling better to the amount of spectral information of each source and therefore being able to better exploit it.

As an example, in Fig. \ref{fig:example} we can see one of the test acoustic scenes. We can see how the output of the icoCNN had a high number of identity switches even when only one source was active but the PI-RNN was able to fix these switches and also reduce the localization error.

\subsection{Attention matrices}

We can interpret the attention matrix of the multi-head attention module of the PI-RNN as an assignment matrix where each row indicates which elements of the input and state set were employed to compute each element of the output set. 

\begin{figure}[ht!]
    \centering
    \subfloat[\label{fig:matrixI}]{
        \includegraphics[width=0.95\linewidth, trim={0 1.5cm 0 1.75cm}, clip]{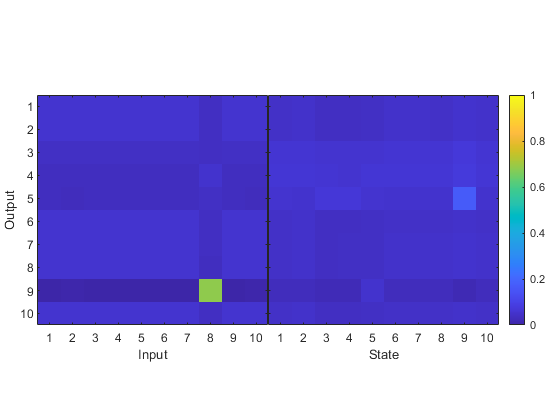}
    } \hfill
    \subfloat[\label{fig:matrixII}]{
        \includegraphics[width=0.95\linewidth, trim={0 1.5cm 0 2.25cm}, clip]{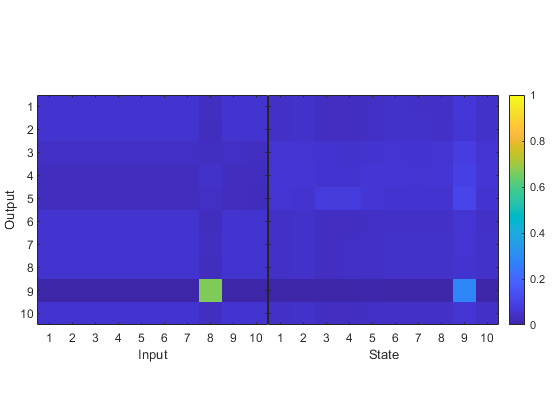}
    } \hfill
    \subfloat[\label{fig:matrixIII}]{
        \includegraphics[width=0.95\linewidth, trim={0 1.5cm 0 2.25cm}, clip]{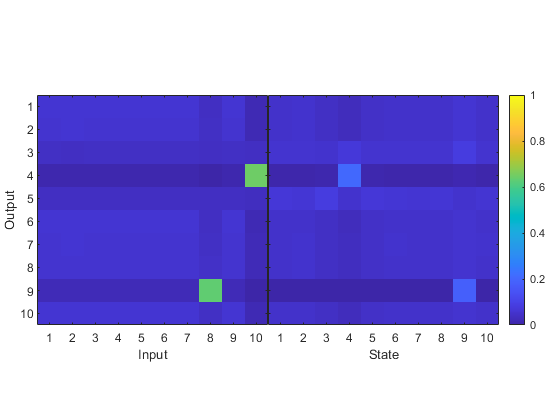}
    } \hfill
    \subfloat[\label{fig:matrixIV}]{
        \includegraphics[width=0.95\linewidth, trim={0 1.5cm 0 2.25cm}, clip]{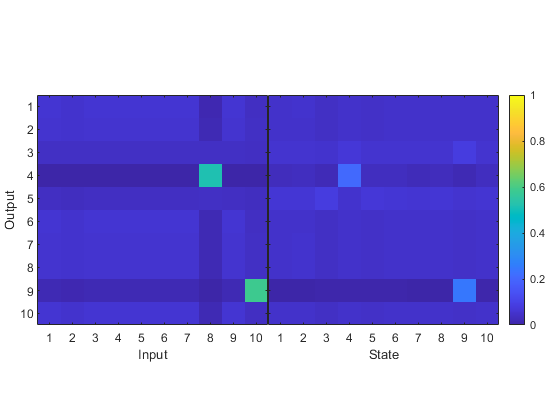}
    } \hfill
    \caption{Examples of attention matrices from the multi-head attention module of the PI-RNN.}
    \label{fig:matrices}
\end{figure}

The attention matrix shown in Fig. \ref{fig:matrixI} corresponds to the first frame where a source appeared and we can see how it was detected at the \nth{8} output of the icoCNN (i.e., the \nth{8} input of the PI-RNN) and the PI-RNN assigned it to its \nth{9} output. In the attention matrix of the next time frame (Fig. \ref{fig:matrixII}) we can see that the \nth{9} output of the PI-RNN was computed combining the information of the new estimate at that frame with the corresponding recurrent state. A new source was detected by the icoCNN at its \nth{4} output in the time frame corresponding to Fig. \ref{fig:matrixIII} and it was assigned to the \nth{10} output of the PI-RNN. Finally, in Fig. \ref{fig:matrixIV} we can see how, after an identity switch at the output of the icoCNN, the PI-RNN was able to assign every new estimate to the correct tracked trajectory fixing the identity switch.

\section{Conclusions}

We have presented a new RNN architecture whose input and state are presented with sets instead of vectors and that is invariant to the permutation of the elements of the input and equivariant to the permutations of the elements of the state set. This new architecture is able to exploit the permutation symmetries of the tracking problem and to outperform the conventional RNN in the preliminary experiments presented in this paper. We expect the difference between the performance of the PI-RNNs and the conventional RNNs to become even greater when including more information of every source at their input.

\bibliography{MicrophoneArraysRedux.bib}

%
%
%

\end{document}